\documentstyle[aas2pp4,psfig]{article}

\slugcomment{To appear in ApJ number NN}

\lefthead{R.J.\ Allen et al.}
\righthead{Large-Scale Dissociation of Molecular Gas in M81}

\newcommand{\Av}{A$_{\rm V}$}
\newcommand{\HII}{\mbox{H\,{\sc ii}}}
\newcommand{\CII}{\mbox{[C\,{\sc ii}]}}
\newcommand{\HI}{\mbox{H\,{\sc i}}}
\newcommand{\Htwo}{H$_{2}$}

\newcommand{\twCO}{$^{12}$CO}

\newcommand{\inthms}[3]{$#1^{\rm h}#2^{\rm m}#3^{\rm s}$}

\newcommand{\intdms}[3]{$#1^{\circ}#2'#3''$}

\newcommand{\pcmcub}{\mbox{${\rm cm^{-3}}$}}
\newcommand{\kmps}{\mbox{${\rm km\;s^{-1}}$}}
\newcommand{\pcmsq}{\mbox{${\rm cm^{-2}}$}}
\newcommand{\pwr}[2]{\mbox{$#1 \times 10^{#2}$}}

\newcommand{\Ico}{I$_{\rm CO}$}

\newcommand{\Hasurf}{erg$\,$cm$^{-2}\,$s$^{-1}\,$arcsec$^{-2}$}
\newcommand{\UVsurf}{erg$\,$cm$^{-2}\,$s$^{-1}\,$\AA$^{-1}\,$arcsec$^{-2}$}
\newcommand{\UVflux}{erg$\,$cm$^{-2}\,$s$^{-1}\,$\AA$^{-1}$}
\newcommand{\lsim}{\mbox{$\mathrel{\vcenter{\hbox{\ooalign{\raise3pt\hbox{$<$}\crcr \lower3pt\hbox{$\sim$}}}}}$}}
\newcommand{\gsim}{\mbox{$\mathrel{\vcenter{\hbox{\ooalign{\raise3pt\hbox{$>$}\crcr \lower3pt\hbox{$\sim$}}}}}$}}

\begin{document}


\title{Evidence for the Large-Scale Dissociation of Molecular Gas\\
in the Inner Spiral Arms of M81}

\author{R.\ J.\ Allen}
\affil{Space Telescope Science Institute, 3700 San Martin Drive\\
Baltimore, MD 21218, USA (rjallen@stsci.edu)}

\author{J.\ H.\ Knapen}
\affil{Department of Physical Sciences, University of Hertfordshire\\
College Lane, Hatfield, Herts AL10 9AB, UK (knapen@star.herts.ac.uk)},

\author{R.\ Bohlin}
\affil{Space Telescope Science Institute, 3700 San Martin Drive\\
Baltimore, MD 21218, USA (bohlin@stsci.edu)}

\and

\author{T.\ P.\ Stecher}
\affil{Code 680, Goddard Space Flight Center\\
Greenbelt, MD 20771 (stecher@uit.gsfc.nasa.gov)}


\begin{abstract}
We compare the detailed distributions of \HI, H$\alpha$, and 150
nm far-UV continuum emission in the spiral arms of M81 at a resolution
of $9''$ (linear resolution 150 pc at 3.7 Mpc distance).  The bright
H$\alpha$ emission peaks are always associated with peaks in the far-UV
emission.  The converse is not always true; there are many regions of
far-UV emission with little corresponding H$\alpha$. The \HI\ and the
far-UV are always closely associated, in the sense that
the \HI\ is often brightest around the edges of the far-UV emission. 

The effects of extinction on the morphology are small, even in the
far-UV. Extensive far-UV emission, often with little corresponding
H$\alpha$, indicates the presence of many ``B-stars'', which produce
mostly non-ionizing UV photons.  These far-UV photons dissociate a
small fraction of an extensive layer of \Htwo\ into \HI.

The observed morphology can be understood if ``chimneys'' are common in
the spiral arms of M81, where holes are blown out of the galactic disk,
exposing the bright \HII\ regions and the corresponding far-UV
associated with vigorous star formation.  These ``naked'' star-forming
regions show little obscuration.  \Htwo\ is turned into \HI\ by UV
photons impinging on the interior surfaces of these chimneys.

The intensity of the far-UV radiation measured by UIT can dissociate
the underlying \Htwo\ with a typical density of $\sim 10$ H nucleii
\pcmcub\ to produce the observed amount of \HI\ in the spiral arms of
M81.  Except for thin surface layers locally heated in these
photo-dissociation regions close to the far-UV sources, the bulk of the
molecular gas in the inner disk of M81 is apparently too cold to
produce much \twCO(1-0) emission.
\end{abstract}


\vspace{3cm}

Accepted for publication in the Astrophysical Journal


\setcounter{footnote}{0}	

\section{Introduction}\label{introduction}

The possibility that an appreciable fraction of the atomic hydrogen
(\HI) found in the spiral arms of disk galaxies is dissociated
molecular gas was first suggested about a decade ago from a comparison
of the relative placement of the dust lanes, \HII\ regions (H$\alpha$),
and \HI\ ridges in parts of M83 (Allen, Atherton, \& Tilanus 1985,
1986) and M51 (Tilanus \& Allen 1987).  This interpretation of the
observations is based on the density-wave model for spiral structure,
in which the temporal sequence of massive star formation is spread out
in space across a spiral arm by the action of a density wave driving
the interstellar gas.  More detailed work has subsequently been done,
including comparisons with other spiral tracers, such as the nonthermal
radio continuum and the CO emission, for M83 (Tilanus \& Allen
1993), for M51 (Vogel, Kulkarni, \& Scoville 1988; Tilanus \& Allen
1989, 1990, 1991; Rand, Kulkarni, \& Rice 1992; Knapen et al. 1992), and
more recently for M100 (Knapen \& Beckman 1994, 1996).

Shaya \& Federman (1987) provided the first theoretical discussion of
the UV photodissociation process as an explanation for the
generally-flat radial distribution of \HI\ in galaxies.  More recently,
from observations of the \CII\ emission from galaxies and quantitative
modelling of the results, Stacey et al.\ (1991) have confirmed that
photodissociation regions (PDRs; e.g.\ Hollenbach \& Tielens 1995) are
the sites of substantial \HI\ production in galaxy disks.
Madden et al.\ (1993) have even suggested that, in their preferred model,
virtually all of the \HI\ in the disk of NGC 6946 could be produced in
PDRs. Unfortunately, the angular resolution available with
current instrumentation in the \CII\ line at $158 \mu$ is inadequate
for the study of individual PDRs in nearby galaxies.

The morphological details of tracers in spiral arms represent new
evidence for the production of \HI\ by the extensive dissociation of
\Htwo\ in PDRs. There are two new elements in the present study. First,
M81 (NGC3031) is a galaxy which is not bright in CO emission (e.g.\ Brouillet
et al.\  1991). The conventional interpretation using CO as a tracer for
\Htwo\ is that this galaxy contains relatively little molecular gas,
contrary to the situation for the previously-studied galaxies M83, M51,
and M100.  Second, the availability of UV image data for M81 from the
Ultraviolet Imaging Telescope (UIT) permits a comparison of the
\HI\ morphology to the distribution of UV photons at $\lambda \approx
150$ nm ($\approx 8.3$ eV), in the middle of the range of photon
energies which are important for the dissociation of \Htwo\ (Stecher \&
Williams 1967)\footnote{Photons with energies in the range 11.2 to 13.6
eV are required to start the process, by pumping the \Htwo\ molecules
into higher vibrational states of the ground electronic state (through
Lyman-Werner band transitions). Subsequently, photons with lower
energies down to $\sim 6.6$ eV can then also dissociate \Htwo\ by
facilitating transitions into the vibrational continuum of the ground
electronic state.}. To interpret the UV data, any effects of patchy
obscuration by dust must be understood to properly compare the UV and
\HI\ morphologies. H$\alpha$ images of M81 aid in our investigation of
this effect.  In our Galaxy, the far-UV extinction (in mags) at
$\lambda \approx 150$ nm is greater than the extinction at H$\alpha$
($\lambda \approx 650$ nm) by a factor of more than 3. Therefore, any
effect of varying obscuration on the morphology of the H$\alpha$ in M81
is greatly magnified in the far UV.

\section{Observations} \label{observations}

The 640 sec far-UV exposure of M81 (FUV0556; Hill et al.\  1995) is from the
UIT/ASTRO-1 archive, and is centered at $\lambda = 152$ nm with bandwidth
35.4 nm (Stecher et al.\  1992).  The VLA \HI\ data were taken by B.\ Hine
\& A.\ Rots, and are provided by D.\ Westpfahl\footnote{New VLA
\HI\ data on M81 have recently been published by Adler \& Westpfahl
(1996). The sensitivity of the new data is significantly better than
that of our older VLA data. However, we are primarily
interested in the bright features on the \HI\ image, and these are not
noise limited in the older data.  Furthermore, the spatial resolution
of our data is better ($9''$ {\it vs.} $12''$) and more
suitable for comparison with the UV image.}.  We used the \HI\ data
cube to calculate a map of \HI\ column density N(\HI).  Only those
pixels with absolute values in excess of 2.5 times the rms noise in the
channel maps are used in the moment calculation, with the additional
restriction that such values must occur in at least two adjacent
velocity channels (which are separated by 10 \kmps). M.\ Kaufman and
N.\ Devereux provided H$\alpha$ images of M81. Although we use the
newer data (Devereux, Jacoby, \& Ciardullo 1995) for our illustrations,
the older data (Kaufman et al.\  1989a) provide an important ``second
opinion'' in a few cases where smoothed cosmic-ray hits on the CCD look
like \HII\ regions, or where foreground stars are imperfectly
subtracted.

In combining the different data sets, we first ensure that the
astrometry of the individual images is consistent by checking the
positions of the center of the galaxy and several foreground stars (UV
and H$\alpha$) and the location and orientation of the general spiral
arm features (\HI).  The original astrometry provided with the data
sets is excellent. Any difference in alignment between images appears
to be no greater than $1'' - 2''$, which is negligible in comparison
with the final resolution of $9''$ used here.

The next step is to smooth the different images to a common
resolution of $9''$, which is that of the \HI\ image (linear
resolution 150 pc at our adopted distance of 3.7 Mpc).  The angular
resolutions of the UV and H$\alpha$ images are determined by fitting 2D
Gaussian profiles to a number of point-like sources in the frames.  We
then convolve the images with the appropriate Gaussians in order to
obtain smoothed images at the resolution of the \HI\ image. The images are
then interpolated to a common pixel grid in order to facilitate a
direct comparison of the positions of various features.

\section{Results and Analysis} \label{results}

Full-sized UIT images of M81 in the near- and far-UV (FUV) are shown as
Figures 1 \& 2 in Hill et al.\  (1995) and are not duplicated here.
Similarly, the full H$\alpha$ image is Figure 1 in Devereux
et al.\ (1995). Our independent derivation of the \HI\ column density map
is essentially identical to Figure 13 in Kaufman et al.\ (1989a).  For
reference, Figure \ref{fullgal} shows a contour representation of the
FUV image (smoothed to $9''$), overlaid on a grey-scale representation
of the \HI\ image of the galaxy. Five frames indicate the locations of
the sub-fields, which we will discuss in more detail below. First, the
radial distributions of the three tracers are considered.

\subsection{Radial distributions}

Figure \ref{radials} shows the radial distributions of the three
tracers: \HI\ (bottom panel), FUV (middle panel), and H$\alpha$ (top
panel).  The curves are derived from the input images by averaging in
elliptical annuli of widths $20''$ spaced every $20''$ ($\approx 330$
pc) along the major axis. We used $i=59^{\circ}$ for the inclination
angle of the galaxy, and PA $=157^{\circ}$ for the position angle of
the major axis (de Vaucouleurs et al.\  1991); these values were kept
constant for all ellipses and for each input image data set.

The large-scale axially-symmetric morphology of the galaxy is reproduced
clearly in the profiles of Fig.\ref{radials}.  Both the FUV and
H$\alpha$ are centrally peaked, whereas the \HI\ profile reflects the
central deficiency of atomic hydrogen seen in Fig.\ref{fullgal}.
From $3'$ to $11'$ the three tracers show similar features: First, the
FUV and H$\alpha$, which trace massive star formation, both rise to
their maximum values near $R\approx 5'$, after which they both
generally decline. More surprising is the detailed correlation of the
peaks and valleys of these two profiles between $5'$ and $10'$ radius,
which strongly suggests that dust extinction does not appreciably
affect the morphology of either the FUV or the H$\alpha$ emission.  In
the following section we show that even the details of the FUV and
H$\alpha$ images on the $\approx 150$ pc scale appear to be largely
unaffected by dust. Second, the \HI\ distribution shows roughly the
same behavior as the massive-star tracers.  The steep rise between $3'$
and $5'$ discussed above is clearly reproduced in \HI, as are the
relative dip around $R\approx 7'$ and the secondary peak at $R\approx
8.5'$. The conventional interpretation of this behavior is simply that
more \HI\ leads to more star formation. However, our preferred
alternative interpretation is that the \HI\ radial profile naturally
follows those of FUV and H$\alpha$ because the \HI\ is a {\em product}
of the star formation, through dissociation, rather than a precursor to
it.  Thus, where the star formation is enhanced, more \HI\ is produced,
which is just what the radial profiles show.

\subsection{Detailed morphology}

Earlier work on M81 (e.g.\ Kaufman et al.\  1989a) and on many other
nearby galaxies shows that the \HI\ and the H$\alpha$ trace spiral
structure in much the same general way at a linear resolution of $\sim
1$ kpc. Hill et al.\  (1995) also showed the excellent correspondence
between H$\alpha$ peaks and FUV peaks in M81. The \HI-FUV images
overlaid in Fig.\ref{fullgal} further illustrate this general
coincidence.  However the detailed correspondence between FUV,
H$\alpha$, and \HI\ emission is more complicated.  To illustrate this
richness of detail, we present and discuss the images in 5 fields
positioned as indicated in Fig.\ref{fullgal}.  In
Figs.\ \ref{fieldone}-\ref{fieldfive}, we show for each field first the
contours of the FUV emission overlaid on a grey-scale representation of
itself, then overlaid on the H$\alpha$, and finally on the
\HI\ emission.  In each middle panel we also mark the $B$-band dust
filaments of Kaufman, Elmegreen \& Bash (1989b).

\subsubsection{Far UV -- H$\alpha$ comparison:}

Dust in the Galaxy is unevenly distributed, and the continuum
extinction at $\lambda \approx 150$ nm is about 4 times that at visual
wavelengths. Therefore the FUV image ought to show significantly more
small-scale structure than the H$\alpha$ image.  Instead, the opposite
occurs; the FUV emission is smoother and spread more widely than the
H$\alpha$, as can be seen from a comparison of the ``FUV-on-FUV'' and
``FUV-on-H$\alpha$'' panels in Figs.\ \ref{fieldone}-\ref{fieldfive}.

There are two aspects to this unexpected H$\alpha$-FUV correspondence.
First, every reliable bright peak in the H$\alpha$ has a counterpart in
the FUV. Second, there are many places with extended and relatively
intense FUV emission which have little or no associated H$\alpha$.

The fact that every reliable bright peak in the H$\alpha$ has a
counterpart in the FUV must mean that the effects of extinction on the
morphology are small on the scale of 150 pc, so that either both the
H$\alpha$ and the FUV appear either essentially unobscured, or both are
partly hidden in approximately the same way. The extinction may also be
so high that {\em neither} the FUV {\em nor} the H$\alpha$ are
visible\footnote{The regions of highest dust extinction appear to be
filamentary, as indicated for example on the central panel of
Figs.\ref{fieldone}-\ref{fieldfive} in this paper. With such a patchy
geometry it is unlikely that {\em all parts} of a large \HII\ region
consisting of half a dozen O stars could be hidden, as is corroborated
by detailed comparisons of the thermal radio emission with the
H$\alpha$ in e.g.\ M101 (Israel, Goss, \& Allen 1975), NGC 6946 (van
der Kruit, Allen, \& Rots 1977), and M51 (Tilanus et al.\  1988). None of
those galaxies appear to harbor large \HII\ regions which are entirely
hidden by dust.}.

There are a few regions which appear to emit relatively
strong H$\alpha$ but do not have a counterpart in FUV.  Such regions
are for instance located at
\inthms{09}{52}{12}, \intdms{69}{18}{15} (in the middle
of a UV void in Field~1), at
\inthms{09}{50}{33}, \intdms{69}{17}{30} (linear structure
$\sim 30''$ long in Field~3), or at
\inthms{09}{50}{36}, \intdms{69}{20}{10} (small crescent-shaped
object in Field~3). 
However, none of these H$\alpha$ features appear on the older H$\alpha$
image of Kaufman et al.\  (1989a) (not shown here), confirming that these
features on the Devereux et al.\  H$\alpha$ image are most probably not
\HII\ regions at all, but artifacts such as (smoothed) cosmic ray
events or improperly-subtracted foreground stars\footnote{The older
H$\alpha$ image of Kaufman et al.\ \ also shows a number of
``\HII\ regions'' which do not appear on the Devereux et al.\  H$\alpha$
image.}.

While every reliable bright peak on the H$\alpha$ image has a
counterpart in the FUV image with the same position and general
shape, there are many places with extended
and relatively intense FUV emission which have little or no
associated H$\alpha$. There are even regions of relatively bright
FUV emission, such as those at
\inthms{09}{52}{10}, \intdms{69}{16}{30} (Field~1), and
\inthms{09}{51}{03}, \intdms{69}{23}{30} (Field~4), which show hardly
any counterparts in H$\alpha$ at the present levels of sensitivity.

In principle, some of these UV sources could be artifacts on the UIT FUV
image. There are five kinds of artifacts in the UIT data which could
be important for us: 1) {\em ``Bright Spots'':} There are one or two bright
spots in each UIT camera; they occur within 200 pixels of the edge of
the image. The M81 image is in the central parts of the field, so we
are not concerned with these artifacts. 2) {\em Cosmic Rays:} These are
rare on UIT images, ranging from none to several per image. They are
always very bright, saturated, and extended blobs which do not look like
stars. The full-resolution UIT FUV frame we have used does not appear
to have such features. 3) {\em Foreground Galactic Stars:} Typically
there are one or two UV-bright Galactic foreground stars in each UIT
field at the intermediate Galactic latitudes of M81. These are mostly V
$\approx 11$ mag A stars, so they are bright in optical CCD images.
Again, there are no such stars known within the optical image of M81.
4) {\em Scratches:} The UIT detector is photographic film, and scratches can
appear on it from the handling and developing processes. These are visible
at faint levels, but are easily recognizable as long narrow parallel
lines. Smoothing will reduce their impact, and our M81 image does not
show the effects of scratches. 5) {\em Dust Specks:} During densitometry
of the flight film in a clean room environment, some dust contamination
still appears, which mimics stars. Since only the larger specks are
important for this paper, the reality of all bright sources is easily
confirmed by comparison with the UIT 128 sec shorter exposure of the
same field (FUV0557).

The features in M81 which are relatively strong in the
UV but weak in H$\alpha$ are not artifacts. Apparently, the
stars which power these regions produce ample amounts of dissociating
FUV photons at $\lambda \approx 150$ nm, but relatively few ionizing
photons at $\lambda \leq 91.2$ nm. We shall refer here to this stellar
population as ``B-stars'', but recognize that late O and early A stars
may be included. Differences in the relative amounts of UV and
H$\alpha$ in any given region could arise from differences in age, as
has been discussed for M81 by Hill et al.\ (1995), and/or from variations
in the IMF.

\subsubsection{Far UV -- \HI\ comparison:}

From the third panel in
Figs.\ref{fieldone}-\ref{fieldfive}, \HI\ is often found alone at faint
levels in the spiral arms, without closely-associated UV and with
little or no H$\alpha$ at the present sensitivity.  However,
in the neighborhood of a FUV concentration, the \HI\ tends to be
brighter, and the brightest peaks are located at the {\em periphery} of
the UV concentrations rather than at the position where the UV has its
local maximum. The \HI\ may show a peak slightly offset from the
FUV peak, for example in the small UV region near
\inthms{09}{52}{02}, \intdms{69}{19}{30} (Field~1);
but, more often, the \HI\ forms an elongated patch next to the FUV, such
as near
\inthms{09}{52}{19}, \intdms{69}{12}{30} (Field~2), or
\inthms{09}{50}{31}, \intdms{69}{20}{45} (Field~3).
On a larger scale, \HI\ ridges a kiloparsec ($\sim\,1'$) or more
in length are found preferentially alongside UV ridges, such as the
U-shaped \HI\ feature stretching across Field~2 along the northern edge
of the UV near
\inthms{09}{52}{00}, \intdms{69}{13}{20},
and the \HI\ ridge stretching in a N-S direction just to the east of
the UV in Field~3 near
\inthms{09}{50}{40}, \intdms{69}{19}{00}.

Some of the most interesting cases are those where the \HI\ surrounds
the UV concentration, such as the FUV peak sitting inside the \HI\
depression near
\inthms{09}{50}{39}, \intdms{69}{18}{45} (Field~3), and another at\\
\inthms{09}{52}{14}, \intdms{69}{18}{40} (Field~1).
A rather spectacular example of this morphology is the region near
\inthms{9}{50}{44}, \intdms{69}{17}{30} (Field~3)
where the \HI\ forms a ``halo'' surrounding two peaks of a large
UV-emitting region. The northern peak of this pair corresponds with the
relatively bright \HII\ region \#44 in the catalog of Petit, Sivan, \&
Karachentsev (1988) and is also \#131 in Kaufman et al.\ \ (1987). The
southern peak is actually a loose group of small, faint \HII\ regions
including \#55 in Petit et al.\ \ and several others, which also
appear in Fig.3 of Kaufman et al.\  (1987) but are apparently too faint
to be catalogued there.

\subsection{Statistics of the associations}

\subsubsection{H$\alpha$ associated with FUV peaks:}

In the five fields we have identified a total of 144 sites of
locally-enhanced FUV emission (``FUV peaks''), of which 21 are
``strong'' \\
(\pwr{\geq 10}{-18} \UVsurf).
In 63\% of the cases (90), H$\alpha$ emission
is associated and coincident with the FUV peak, in 13\% (18) the
associated H$\alpha$ emission is slightly offset. For 10\% (14 peaks)
of the FUV peaks no associated H$\alpha$ emission is evident on our
images at the present level of sensitivity. Another 15\% (22) could not
be easily classified; this category includes regions with diffuse
H$\alpha$ emission, possible unsubtracted residuals from foreground
stars in the H$\alpha$ image, and sources on the UV image which may be
Galactic foreground stars.

\subsubsection{\HI\ associated with FUV peaks:}

Of the 144 FUV peaks, only 8\% (11 peaks) do not show any appreciable
\HI\ emission that can reasonably be considered as associated with
the UV peak (i.e.\ within $\sim 200$ pc). In 11\% of the cases (16) the \HI\
is coincident with the FUV, but in a striking majority of 81\% of all
FUV peaks the \HI\ is seen offset from the FUV.  The latter category
includes cases where the bright \HI\ peak is offset from the FUV, but
there is still emission from the wings of the \HI\ extending towards
the position of the FUV peak.  Of our 21 strong FUV peaks, one
(5\%) may have coincident \HI\ emission, but 20, or possibly all
(100\%) have associated \HI\ emission offset from the FUV peak.
 
\section{Discussion}\label{discussion}

The conventional description of the star formation process begins with
the (possibly triggered) gravitational collapse of massive gas clouds
in the disk of a galaxy. These ``super-clouds'' may be in the form of
\HI\ or \Htwo, depending on the physical conditions in the ISM
(e.g.\ Elmegreen 1991 and references therein; Elmegreen 1993).  The
denser parts of these clouds cool further (and turn into \Htwo\ if they
aren't already in that form) in giant molecular clouds (GMCs); stars
then form and evolve in these GMCs (e.g.\ Shu et al.\  1993). In this
picture, a small fraction of the gas is consumed in the star formation
process, and a larger fraction may be ionized and/or evacuated by
stellar winds and supernovae. Some photodissociation of \Htwo\ into
\HI\ has also been observed on the surfaces of specific GMCs in our
Galaxy (e.g.\ Blitz 1993, section 6; Andersson \& Wannier 1993; Kuchar
\& Bania 1993) and column densities of order \pwr{1}{21} \pcmsq\ have
been reported; but the photodissociation process is usually not thought
of as being responsible for producing a substantial fraction of the
Galactic \HI.  However, our model for the morphology of the UV,
H$\alpha$, and \HI\ in M81 leads to the conclusion that a large
fraction, and possibly all, of the \HI\ in the inner spiral arms of
that galaxy is a photodissociation product.

\subsection{A preliminary model for the morphology}

The \HI, H$\alpha$, and FUV structures observed in M81 can be
understood if ``chimneys'', like those proposed by Norman \& Ikeuchi
(1989), are common in the spiral arms of this galaxy.  In this picture,
holes \gsim 100 pc in size are blown out of the galactic disk by
concentrations of high-mass star formation activity, similar to that
discovered above the Galactic \HII\ region IC1805 by Normandeau,
Taylor, \& Dewdney (1996). From our vantage point above the disk of
M81, such concentrations of star-forming activity are seen essentially
free of obscuration, since the hot stars have evacuated or destroyed
most of the intervening dust in M81.  The gas
in M81 is largely in the form of \Htwo\, at least in the main part of
the disk; \HI\ is produced extensively by UV photons from the
young stars impinging on the inner surfaces of these
structures.  The detailed structure of such an \HI\ layer may resemble
the \HI\ morphology of the Galactic star-forming region G216-2.5, which
has been explained as a PDR by Williams \& Maddalena (1996). This
region shows a layer of \HI\ $\approx 50$ pc thick spread over 300 pc
on the outer surface of a molecular cloud; the \HI\ column density is
of order \pwr{0.2}{21} \pcmsq.  The cause of the photodissociation is
thought to be two young stars located $\sim 50$ pc from the
\HI\ layer.

In Fig.\ref{cartoon} we show a simplified cross-sectional sketch of the
various morphologies which may arise in such a picture. The observer
views the galaxy disk from above. Case I illustrates the IC1805
geometry, which forms the basis for our model. A young star cluster has
formed just above the mid-plane in a dense layer of molecular gas. The
\Htwo\ layer is shown in the sketch as a smooth medium but will
have structure. If O stars are present, the
accompanying H$\alpha$ will be roughly coincident with the UV. Since
the star-formation activity has opened a chimney up into the halo, the
observer sees the optical and UV emission with little obscuration, even
though the \Htwo\ layer underneath the bottom of the chimney will
contain dust and may be opaque.  The \HI, produced from the
\Htwo\ by photodissociation, is spread over the inner surface of the
chimney and will appear to the observer to be enhanced at the edges of
the region, owing to the longer lines of sight there. Case II is the
same geometry as Case I, but since the chimney has blown out on the far
side of a more-or-less opaque disk of gas, the observer may measure
nothing more than a patch of \HI\ emission.

Case III in Fig.\ref{cartoon} shows a particularly energetic
star-forming region located very close to the mid-plane of the galaxy.
In this case, chimneys may blow out on both sides of
the disk, leaving a hole right through the galaxy. The photodissociated
\Htwo\ will appear to the observer as an \HI\ ``cocoon'' wrapped around
the boundary of the UV-emitting region, as seen e.g.\ in Field
3.  Case IV is a one-sided version of Case III, formed in a non-uniform
part of the \Htwo\ layer; it is a more energetic example of the model
for G216-2.5 by Williams \& Maddelena (1996).

\subsection{Production of \HI\ by photodissociation}

A simple way to obtain a rough order of magnitude for the \HI\ column
density in a PDR is to consider the case where the dust opacity
controls the depth to which UV penetrates the surfaces of molecular
clouds.  This will occur when self-shielding by \Htwo\ is relatively
unimportant (low volume density).  In this case, the UV penetrates and
dissociates to a depth where the UV opacity $\tau_{\rm UV} \approx 1$.
With the usual values for dust-to-gas ratio appropriate for translucent
lines of sight in the Galaxy, namely, N(H) = \pwr{1.8}{21}\ $\times$
\Av\ atoms \pcmsq (Bohlin, Savage, \& Drake 1978), and using ${\rm
A_{FUV}} = 2.54\ \times$ \Av\ (Seaton 1979), we have ${\rm A_{FUV}} =
1$ at N(H) = \pwr{8}{20} atoms \pcmsq, predominantly in the form of
\HI\. This value is close to the threshold values of N(\HI) found in
the immediate neighborhood of FUV concentrations in the five fields
under discussion in this paper and shows that the dissociation picture
provides the right order of magnitude for the \HI\ column density.

\subsubsection{A quantitative model:}

Many authors have developed detailed models of PDRs including
self-shielding; the review by Hollenbach \& Tielens (1995) lists many
relevant references. In particular, \HI\ production in PDRs has been
considered by Federman, Glassgold, \& Kwan (1979) for equilibrium
models; time-dependent models have been computed by Roger and
Dewdney (1992) and by Bertoldi \& Draine (1996).
Sternberg (1988) has provided a convenient analytic solution to a
simple semi-infinite slab geometry in equilibrium with UV radiation
incident on one side. The solution gives the steady-state \HI\ column
density along a line of sight perpendicular to the face of the slab, as
a function of the FUV flux $\chi$ in units of Draine's (1978)
standard values $\chi_0$ for the ISM near the sun, and the total volume
density of H nucleii $n = n$(\HI) + $2n$(\Htwo) in the gas.  Using the
parameters in this equation adopted by Madden et al.\  (1993) we have:
\begin{equation}
{\rm N(\HI)} = \pwr{5}{20} \times \ln [1+(\chi/\chi_0)(90/n)]
\label{dissociate}
\end{equation}

As a specific example, we apply this model to the \HI\ and FUV
associated with the \HII\ region at \inthms{9}{50}{44},
\intdms{69}{17}{30} (Field~3), which is the region described earlier
where the \HI\ appears to surround the UV and \HII\ concentrations. The
bright \HII\ region associated with the northern UV peak is \#44 in the
catalog of Petit et al.\  The UV flux of this region integrated over a
circular aperture $16''$ in diameter on the full-resolution image is
given by Hill et al.\ \ as f$_{150}$ = \pwr{1.39}{-15} \UVflux; we
obtain a somewhat larger value of \pwr{2.55}{-15} by direct integration
of the contours of our (smoothed) image in Fig.\ref{fieldthree} over
the same area.

This observed flux must be corrected for extinction at 150 nm. There
are two parts to this correction: Galactic foreground extinction, and
extinction in M81 itself. Consistent with our conclusion that the UV
and \HII\ regions are essentially ``naked'', the extinction in M81
itself is taken to be completely ``gray'', caused by similar
obscuration of parts of the emitting region both at UV and H$\alpha$.
This kind of extinction can be revealed by a comparison of the thermal
radio flux density of an \HII\ region with its H$\alpha$ flux. With
this method, Kaufman et al.\ (1987) derive an average extinction (at
$\approx 650$ nm) for \HII\ regions in M81 of A$_{\alpha} \approx 0.86$
mag, of which they estimate $\approx 0.23$ mag is Galactic foreground
extinction. In our model, this foreground extinction is the only part
which will increase at shorter wavelengths.  Taking A$_{150} = 3.1
\times$ A$_{\alpha}$ (Seaton 1979), we estimate that for \HII\ region
\#44, A$_{150} = 3.1 \times 0.23 + 0.63 \approx 1.34$ mag. The
corrected FUV flux at the earth is then f$_{150}'$ = \pwr{8.77}{-15}
\UVflux, corresponding to the UV flux from $\sim 9$ stars of type O9V
at a distance to M81 of 3.7 Mpc.  Normandeau et al.\  (1996) list 9 O
stars as the major power sources in the Galactic region IC1805, which
is our model for the ``chimneys'' in M81.

The following step is to compute the UV flux at the location of the
\HI\ ``blanket''. The distance between the dissociating stars and the
\HI\ near region \#44 in M81 is about 1/3 of our resolution FWHM, or
about 50 pc, which is also similar to the geometry in IC1805 and the
same as the value estimated by Williams \& Maddalena for G216-2.5.
The flux ratio is therefore (\pwr{3.7}{6}/50)$^2$ = \pwr{5.5}{9} for a
distance to M81 of 3.7 Mpc, so the FUV flux incident on the molecular
cloud is $\chi$ = \pwr{8.77}{-15} $\times$ \pwr{5.5}{9} = \pwr{4.8}{-5}
\UVflux.  The value for $\chi_0$ at 150 nm can be read from Fig.1 in
Van Dishoeck \& Black (1988) as \pwr{2}{5}
photons$\,$cm$^{-2}\,$s$^{-1}\,$\AA$^{-1} \times$ \pwr{1.32}{-11}
ergs/photon = \\
\pwr{2.64}{-6} \UVflux so that the ratio $\chi/\chi_0
\approx 18$.

From equation \ref{dissociate} the density $n$ of
the underlying gas can be estimated from the observed FUV flux and the
\HI\ column density of the PDR. The extended regions of \HI\ emission
in close proximity to \HII\ region \#44 have N(\HI) = \pwr{2.5}{21}
\pcmsq.  There are more localised areas reaching a brightness of two or
three times this value, but in our picture these enhancements are the
result of viewing a thick blanket of \HI\ more or less edge-on (cf.
Fig.\ref{cartoon}). Taking G216-2.5 again as a model, the line of sight
may reach 300 pc tangential to a layer of intrinsic thickness of only
50 pc. Accounting for our 150 pc beam, the net enhancement would be of
order (300/50) $\times$ (50/150) = 2, in rough agreement with our
\HI\ observations. Using N(\HI) = \pwr{2.5}{21} \pcmsq and $\chi/\chi_0
= 18$ in equation \ref{dissociate}, we compute the density of the
underlying gas to be $n \approx 10$ \pcmcub. Although the accuracy of
this result is not very high, the value is about the same as that used
by Rand et al.\ (1992) in their ``best'' photodissociation models for
the \HI\ in M51.  For G216-2.5, we calculate from the values in
Williams \& Maddelena that $n$(\HI) $\approx$ N(\HI)/L = 1.5
\pcmcub\ with L = 50 pc, significantly less than the value $n = n$(\HI)
+ 2$n$(\Htwo) = 10 \pcmcub\ derived here\footnote{The gas in the
\HI\ zone of the PDR will be mostly in the form of \HI.}.  However, the
detected \HI\ will have been heated by the photodissociation process to
$\sim 80$ K, and the factor of $\sim 6$ in density is easily understood
if the kinetic temperature of the precursor \Htwo\ is less than a
plausible $\sim 15$ K.

\subsubsection{The molecular content of M81:}

The photodissociation picture provides a reasonable explanation for the
\HI\ morphologies and column densities in the spiral arms of M81. The
underlying gas is of modest density, $\sim 10$ H nucleii \pcmcub, but must
be widespread, since every UV source has photodissociated \HI\ in its
immediate vicinity.  The amount of \Htwo\ present in the spiral arms of
M81 is uncertain, but it is likely to be at least as much as the \HI,
and may be significantly greater. This result contrasts with the
CO(1-0) surveys of M81. For instance, using the usual factor of Bloemen
et al.\ (1986) to convert \Ico\ to N(\Htwo), Brouillet et al.\ (1991)
conclude that the \Htwo\ mass in the annular region $4 \leq {\rm R}
\leq 7$ kpc containing the major spiral arms in M81 is only 20\% of the
\HI\ mass in the same area.  The ratio varies from 10\% to 40\% in
specific regions along the spiral arms as observed with the 12m NRAO
millimeter radio telescope (beam FWHM $\approx 1'$) when the \HI\ is
smoothed to the same resolution (Brouillet et al.\  Table 3).  The
conventional conversion factor linking \twCO(1-0) surface brightness
and \Htwo\ column density must {\em underestimate} the quantity of
molecular gas in the spiral arms of M81 {\em by at least a factor of
five.} A similar discrepancy was noted by Kaufman et al.\ (1989b) in their
study of the dust lanes in M81.

Allen (1996) reviewed the case for using \twCO(1-0) brightnesses to
infer \Htwo\ column densities in galaxy disks and concluded that this
practice is unreliable. Our present result in M81 contributed to
that conclusion.

\section{Conclusions} \label{conclusions}

Our detailed morphological study of the FUV, H$\alpha$, and \HI\ in M81
at a resolution of 150 pc indicates that most of the \HI\ is
UV-dissociated \Htwo, at least in the inner spiral arms of this galaxy.
In this picture, the \HI\ in the spiral arms is a {\em product} of the star
formation process rather than a precursor.

A corollary of this picture is that the disk of M81 must harbor a large
reservoir of molecular gas which has so far escaped detection, e.g.\ in
the \twCO(1-0) line.  Since M81 has a generally low level of heat
production, as evidenced by the faintness of the nonthermal radio
continuum and the generally low surface brightness in H$\alpha$
emission, this molecular gas is likely to be generally cold and
therefore faint in e.g.\ \twCO(1-0) emission. The ``skins'' of these
molecular clouds may be sufficiently warmed by nearby UV sources to be
visible in \twCO(1-0), but this emission will have a small spatial
filling factor.  The underlying cold clouds may be visible in the inner
disk of M81 as broad emission lines of \twCO(1-0) characteristic of
GMCs, but unusually faint and with abnormally low values of the
\twCO(2-1)/\twCO(1-0) ratio. Such molecular clouds have been found in
the inner disk of M31 by Allen \& Lequeux (1993) and by Loinard
et al.\ (1995).


\acknowledgements

We thank D.\ Westpfahl, M.\ Kaufman and N.\ Devereux for
providing the data for this study, and for helpful
discussions on its use and interpretation. We are also grateful
to A.\ Poglitsch for introducing the notion that PDRs may be
ubiquitous in galaxy disks and to D.\ Hollenbach for suggesting
relevant references on PDR models. It is a pleasure to acknowledge
Colin Norman and all our colleagues at STScI and JHU for
stimulating discussions. Laurent Loinard assisted with a critical
reading of the manuscript and provided the drawing for Figure
\ref{cartoon}. JHK thanks the Director of STScI for supporting several
visits to the Institute while this paper was being written, and the
{\it FCAR Qu\'ebec Action Concert\'ee} for financial assistance.


\clearpage

\onecolumn

\clearpage


\figcaption{Contour representation of the far-UV (FUV) image of M81 taken with
the Ultraviolet Imaging Telescope on the ASTRO-1 Mission, overlaid on
a grey-scale representation of the VLA HI map of M81.  The numbered
squares indicate five sub-fields, discussed in detail in the text. The
resolution of both images is $9'' = 150$ pc.  Contour levels for the
FUV are $2.3, 3.1, 4.6, 6.2, 7.7, 9.2, 13.9, 18.5$ and \pwr{36.9}{-18}
\UVsurf.  Grey scales for the HI vary linearly from $1.5$ (light) to
\pwr{7.3}{21} atoms \pcmsq\ (dark).\label{fullgal} }

\figcaption{
Radial emission profiles of \HI\ (bottom panel),
FUV (middle panel),
and H$\alpha$ (top panel) in the disk of M81 on
logarithmic scales.  Units are log of N(\HI) in atoms \pcmsq, log
of FUV surface brightness in \UVsurf, and log of
H$\alpha$ surface brightness in \Hasurf. \label{radials}  
}

\figcaption{
Field 1 in M81, as indicated in Fig.\protect\ref{fullgal} and
discussed in the text.  Contours in all three panels are of the FUV
emission, contour levels as in Fig.\protect\ref{fullgal}.  The contours are
overlaid on grey-scale representations of the FUV (left panel,
grey-scales from 0.8 [light] to \pwr{15}{-18} \UVsurf\ [dark]);
H$\alpha$ (middle panel; greys from 0.5 to \pwr{95}{-17}
\Hasurf; and HI (right panel, grey
levels as in Fig.\protect\ref{fullgal}).
Positions of the dust filaments as measured by
Kaufman, Elmegreen \& Bash (1989) are indicated by the line segments
drawn in the middle panel. \label{fieldone}
}

\figcaption{
As Fig.\protect\ref{fieldone}, now for Field 2. \label{fieldtwo}
}

\figcaption{
As Fig.\protect\ref{fieldone}, now for Field 3. \label{fieldthree}
}

\figcaption{
As Fig.\protect\ref{fieldone}, now for Field 4. \label{fieldfour}
}

\figcaption{
As Fig.\protect\ref{fieldone}, now for Field 5. \label{fieldfive}
}

\figcaption{
Cartoon sketch of how various ``Norman-Ikeuchi'' chimneys
could occur in the disk of a galaxy (Norman \& Ikeuchi 1989, see their
Figs.\ 4 \& 5). The transverse size scale of a large chimney (powered
by a cluster of O stars) may reach a few hundred pc.  In a dense layer
of \Htwo, these structures develop envelopes of \HI\ which partially
surround the star-forming region.  When viewed from a distant vantage
point, such structures may explain the morphology of the UV, \HI, and
H$\alpha$ which we observe in M81. \label{cartoon}
}


\begin{references}

Adler, D.A., \& Westpfahl, D.J.\ 1996, AJ, 111, 735

Allen, R.J.\ 1996, ``Molecular Gas in Spiral Galaxies'', in ``New
Extragalactic Perspectives in the New South
Africa'', ed. D.L.\ Block \& J.M.\ Greenberg (Kluwer; Dordrecht), 50

Allen, R.J., Atherton, P.D., \& Tilanus, R.P.J.\ 1985,
in ``Birth and Evolution of Massive Stars and Stellar Groups,''
ed.\ W.\ Boland \& H.\ Van Woerden (Reidel; Dordrecht), 243

Allen, R.J., Atherton, P.D., \& Tilanus, R.P.J.\ 1986, Nature, 319, 296

Allen, R.J., \& Lequeux, J.\ 1993, ApJ, 410, L15

Andersson, B.-G., \& Wannier, P.\ 1993, ApJ 402, 585

Bertoldi, F., \& Draine, B.T.\ 1996, ApJ, 458, 222

Blitz, L.\ 1993, in ``Protostars \& Planets III'',
ed.\ E.H.\ Levy \& J.I.\ Lunine (Univ.\ Arizona Press; Tucson), 125

Bloemen, J.B., Strong, A.W., Blitz, L., Cohen, R.S., Dame, T.M.,
Grabelsky, D.A., Hermsen, W., Lebrun, F., Meyer-Hasselwander, H.A.,
\& Thaddeus, P.\ 1986, A\&A 154, 25

Bohlin, R.C., Savage, B.D., \& Drake, J.F.\ 1978, ApJ, 224, 132

Brouillet, N, Baudry, A., Combes, F., Kaufman, M., \& Bash, F.\ 1991,
A\&A, 242, 35

De Vaucouleurs, G., de Vaucouleurs, A., Corwin, H.G., Buta, R.J.,
Paturel, G., \& Fouqu\'e, P.\ 1991, ``Third Reference Catalogue of
Bright Galaxies'' (Springer, New York)

Devereux, N.A., Jacoby, G., \& Ciardullo, R.\ 1995, AJ, 110, 1115

Draine, B.T.\ 1978, ApJS, 36, 595

Elmegreen, B.G.\ 1991, in ``The Physics of Star Formation and Early Stellar
Evolution'', ed. C.\ Lada \& N.\ Kylafis (Kluwer), 35

Elmegreen, B.G.\ 1993, in ``Protostars and Planets III'',
ed.\ E.H.\ Levy \& J.I.\ Lunine (Univ.\ Arizona Press; Tucson), 97

Federman, S.R., Glassgold, A.E., \& Kwan, J.\ 1979, ApJ, 227, 466

Hill, J.K.\ et al.\  1995, ApJ, 438, 181

Hollenbach, D.J., \& Tielens, A.G.G.M.\ 1995, in ``The Physics and Chemistry
of Interstellar Molecular Clouds'', ed. G.\ Winnewisser \& G.C.\ Pelz
(Springer-Verlag; Berlin), 164

Israel, F.P., Goss., W.M., \& Allen, R.J.\ 1975, A\&A, 40, 421

Kaufman, M., Bash, F.N., Kennicutt, R.C.Jr., \& Hodge, P.W.\ 1987, ApJ, 319, 61

Kaufman, M., Bash, F.N., Hine, B., Rots, A.H., Elmegreen, D.M.,
Hodge, P.W.\ 1989a, ApJ, 345, 674

Kaufman, M., Elmegreen, D.M., \& Bash, F.N.\ 1989b, ApJ, 345, 697

Knapen, J.H., Beckman, J.E., Cepa, J., van der Hulst, J.M. \&
Rand, R.J.\ 1992, ApJ, 385, L37

Knapen, J.H., \& Beckman, J.E.\ 1994, in ``Physics of Gaseous and
Stellar Discs of Galaxies'', ed.\ I.\  King (A.S.P.\ Conference Series
Vol.\ 66), 329

Knapen, J.H., \& Beckman, J.E.\ 1996, MNRAS, 283, 251

Kuchar, T.A., \& Bania, T.M.\ 1993, ApJ, 414, 664

Loinard, L., Allen, R.J., \& Lequeux, J.\ 1995, A\&A, 301, 68

Madden, S.C., Geis, N., Genzel, R., Herrmann, F., Jackson, J.,
Poglitsch, A., Stacey, G.J., \& Townes, C.H.\ 1993, ApJ, 407, 579

Norman, C.A., Ikeuchi, S.\ 1989, ApJ, 345, 372

Normandeau, M., Taylor, A.R., \& Dewdney, P.E.\ 1996, Nature, 380, 687

Petit, H., Sivan, J.-P., \& Karachentsev, I.D.\ 1988, A\&AS, 74, 475

Rand, R.J., Kulkarni, S.R., \& Rice, W.\ 1992, ApJ, 390, 66

Roger, R.S., \& Dewdney, P.E.\ 1992, ApJ, 385, 536

Seaton, M.J.\ 1979, MNRAS, 187, 73p

Shaya, E.J., \& Federman, S.R.\ 1987, ApJ, 319, 76

Shu, F., Najita, J., Galli, D., Ostriker, E., Lizano, S.\ 1993,
in ``Protostars and Planets III'', ed.\ E.H.\ Levy \& J.I.\ Lunine
(Univ.\ Arizona Press; Tucson), 3

Stacey, G.J., Geis, N., Genzel, R., Lugten, J.B., Poglitsch, A.,
Sternberg, A., \& Townes, C.H.\ 1991, ApJ, 373, 423

Stecher, T.P., \& Williams, D.A.\ 1967, ApJ, 149, L29

Stecher, T.P.\ et al.\  1992, ApJ, 395, L1

Sternberg, A.\ 1988, ApJ, 332, 400

Tilanus, R.P.J., \& Allen, R.J.\ 1987, in ``Star Formation in Galaxies'',
ed. C.J.\ Lonsdale Persson (NASA CP-2466), 309

Tilanus, R.P.J., \& Allen, R.J.\ 1989, ApJ, 339, L57

Tilanus, R.P.J. \& Allen, R.J.\  1990, in ``The Interstellar Medium
in External Galaxies: Summaries of Contributed Papers,'' ed.  D.J.
Hollenbach and H.A. Thronson, Jr. (NASA Conference Publication 3084), 298

Tilanus, R.P.J., \& Allen, R.J.\ 1991, A\&A, 244, 8

Tilanus, R.P.J., \& Allen, R.J.\ 1993, A\&A, 274, 707

Tilanus, R.P.J., Allen, R.J., van der Hulst, J.M., Crane, P.C., \&
Kennicutt, R.C.\ 1988, ApJ, 330, 667

Van der Kruit, P.C., Allen, R.J., \& Rots, A.H.\ 1977, A\&A, 55, 421

Van Dishoeck, E.F., \& Black, J.H.\ 1988, ApJ, 334, 771

Vogel, S.N., Kulkarni, S.R., \& Scoville, N.Z.\ 1988, Nature, 334, 402

Williams, J.P., \& Maddelena, R.J.\ 1996, ApJ, 464, 247

\end{references}
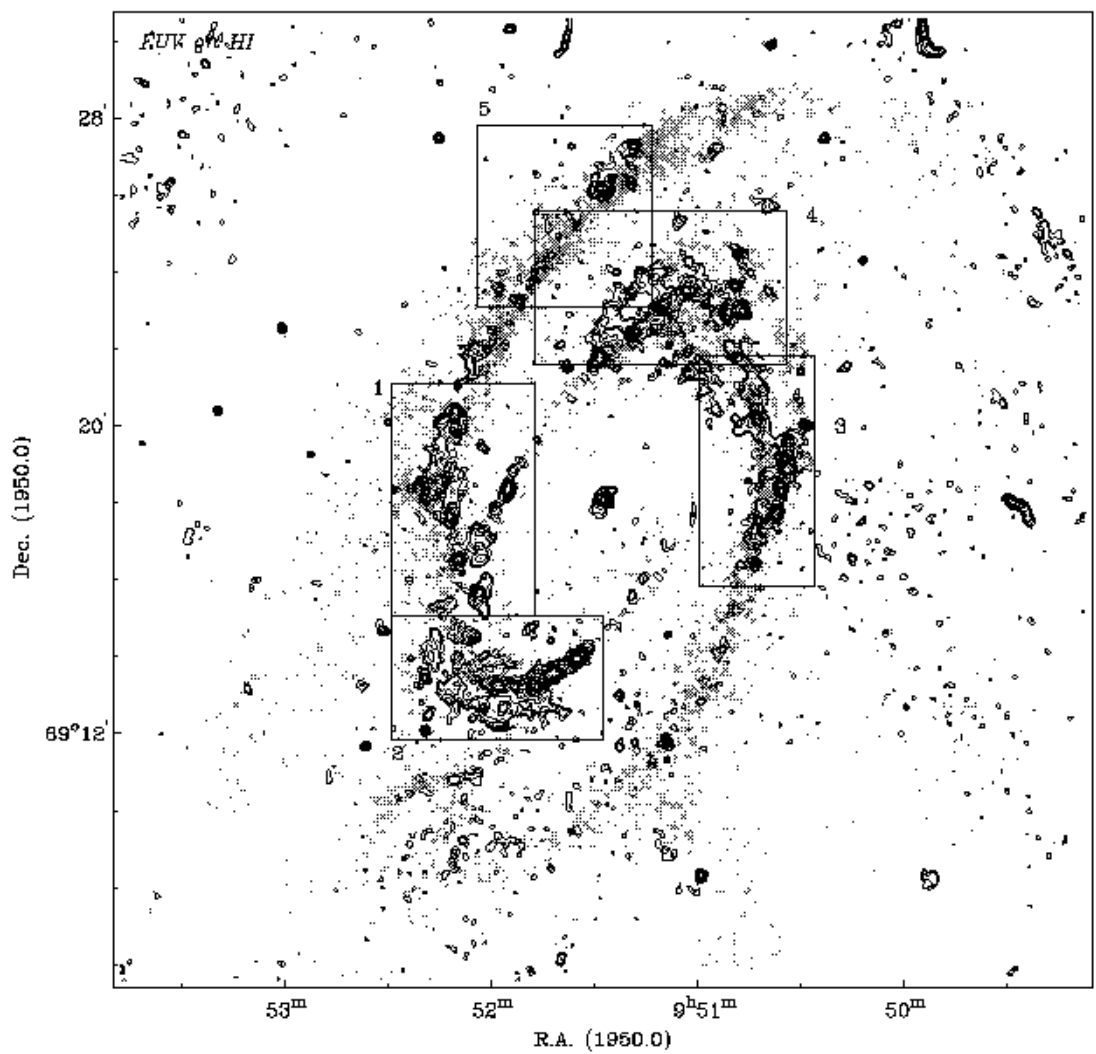
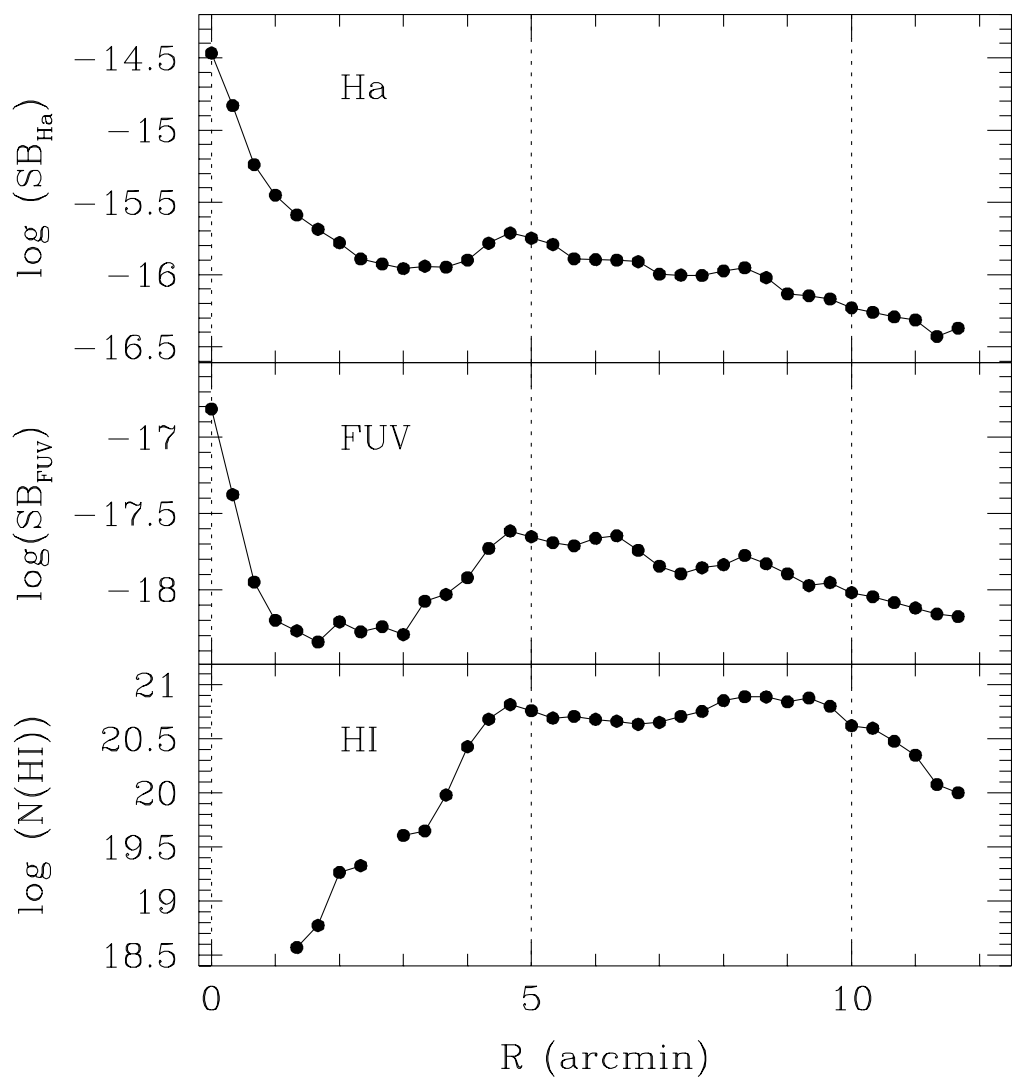
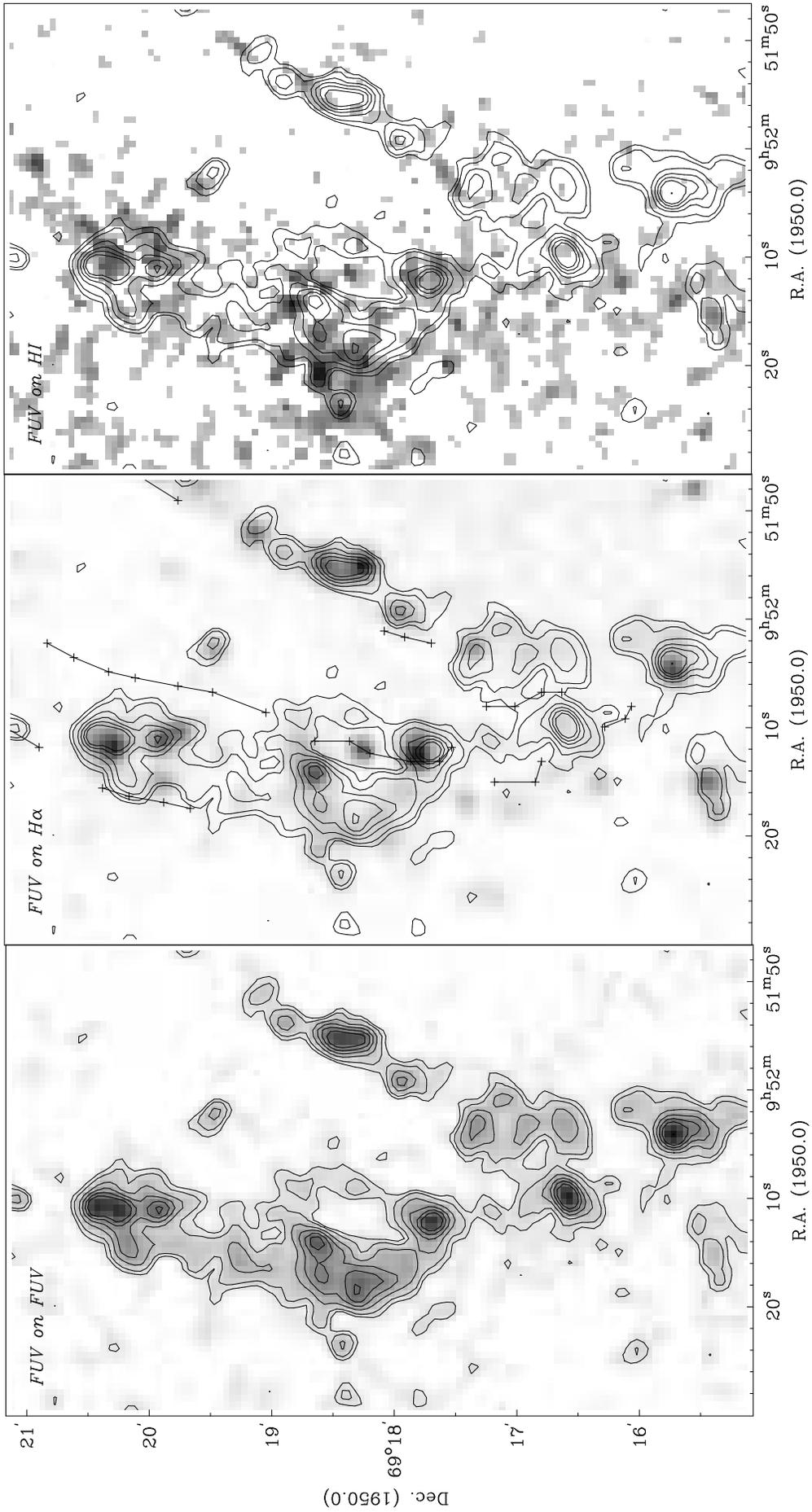
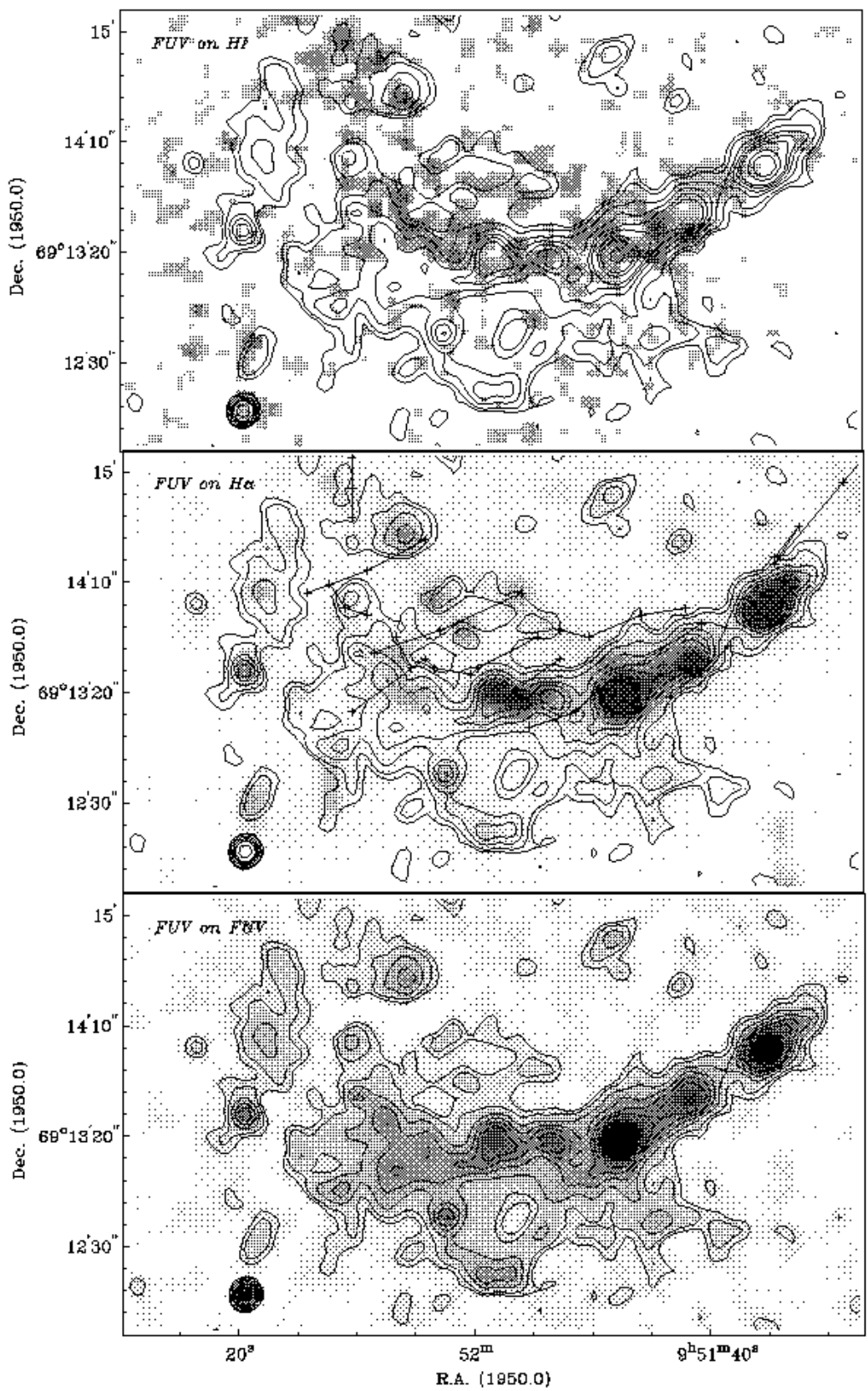
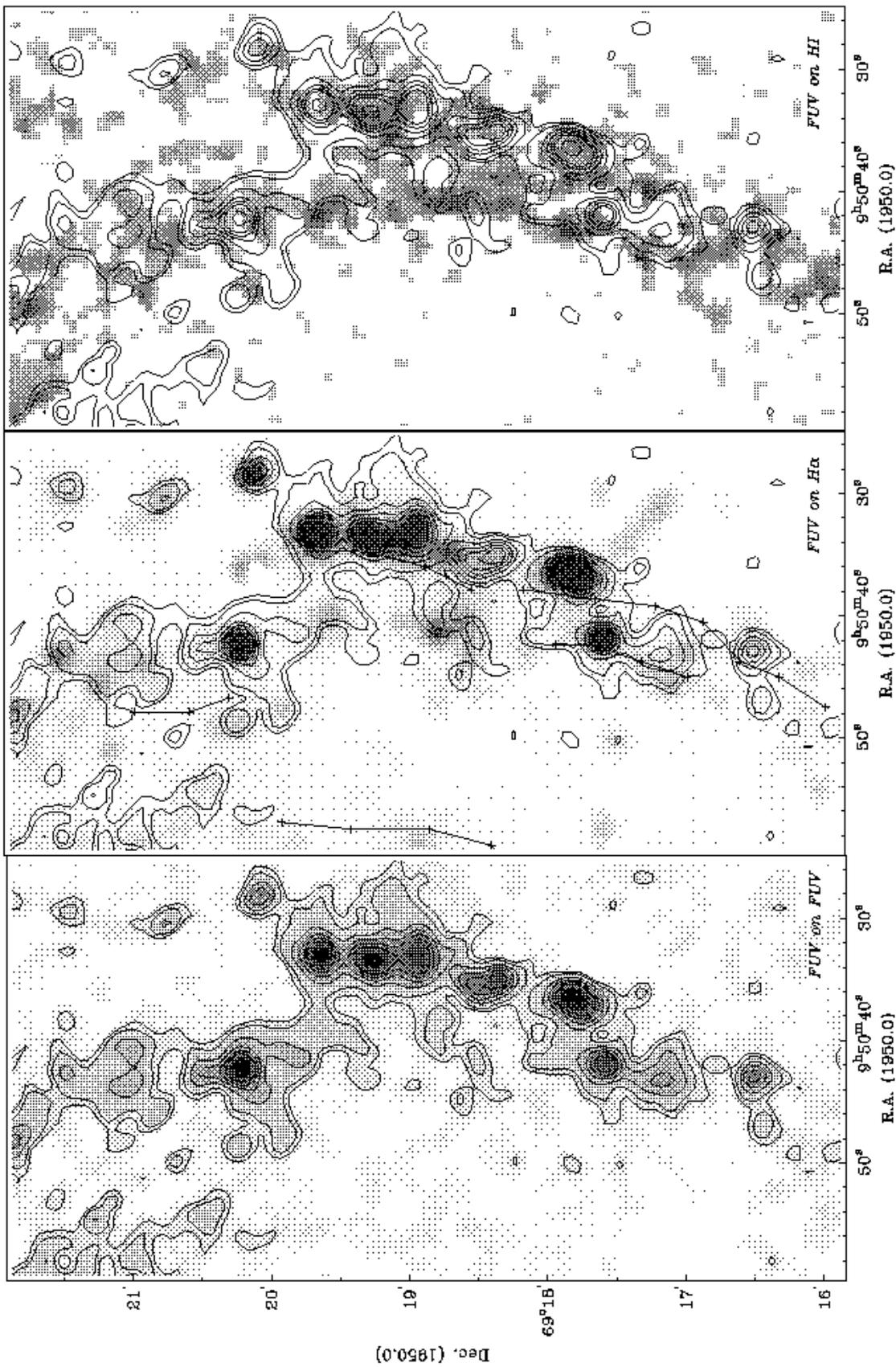
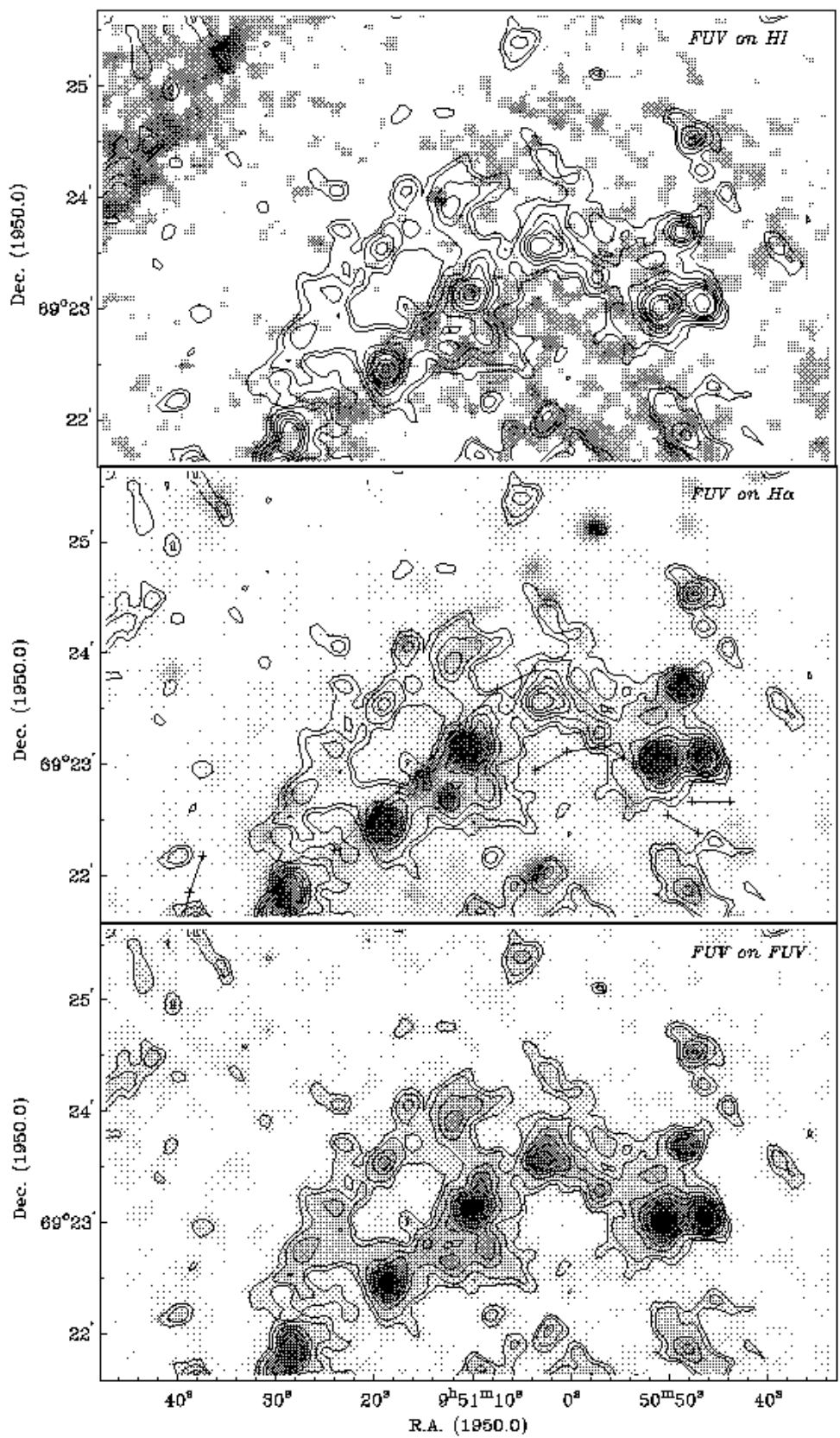
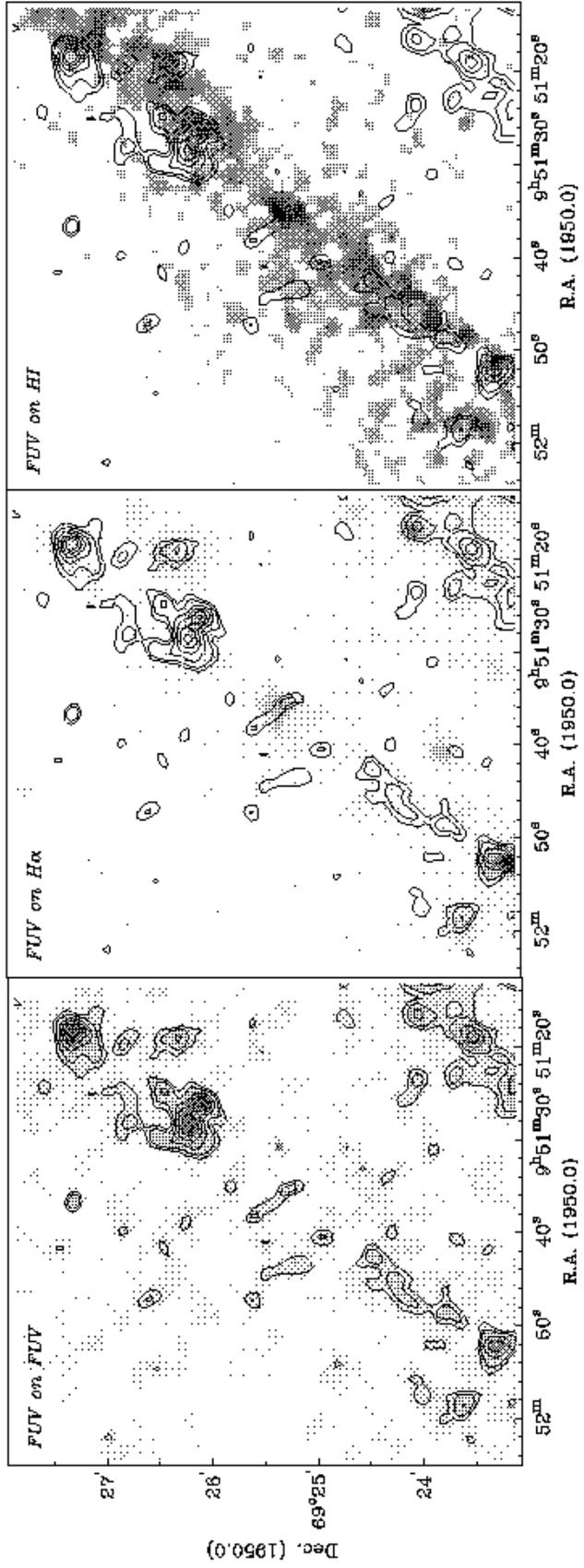
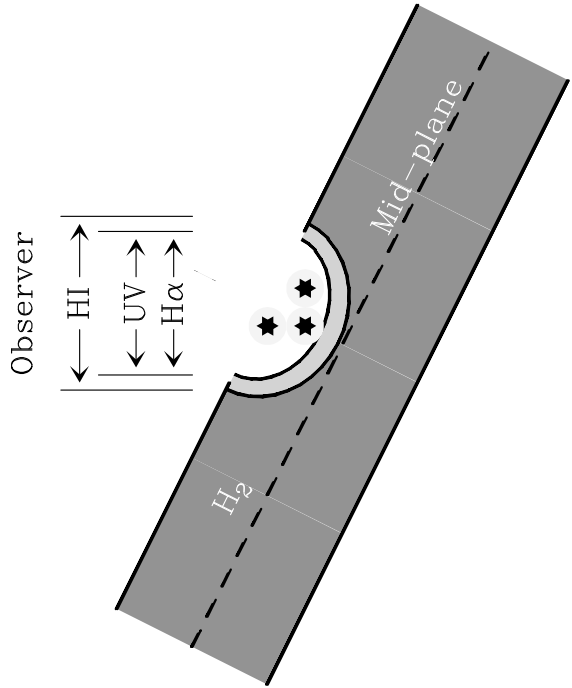
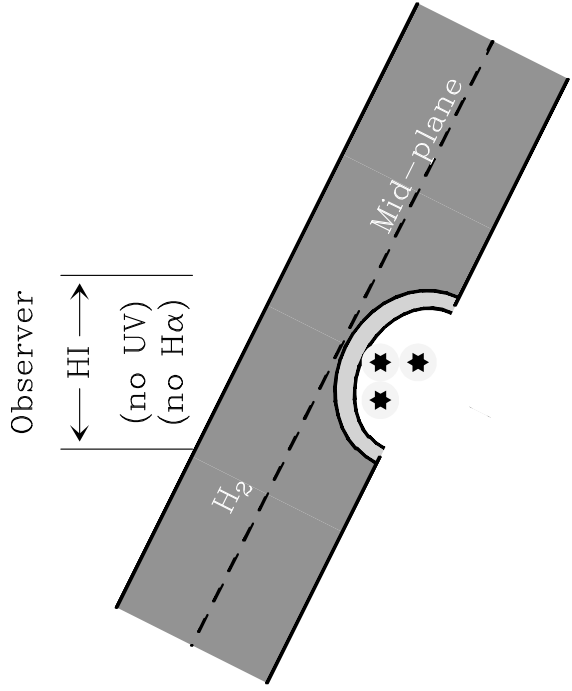
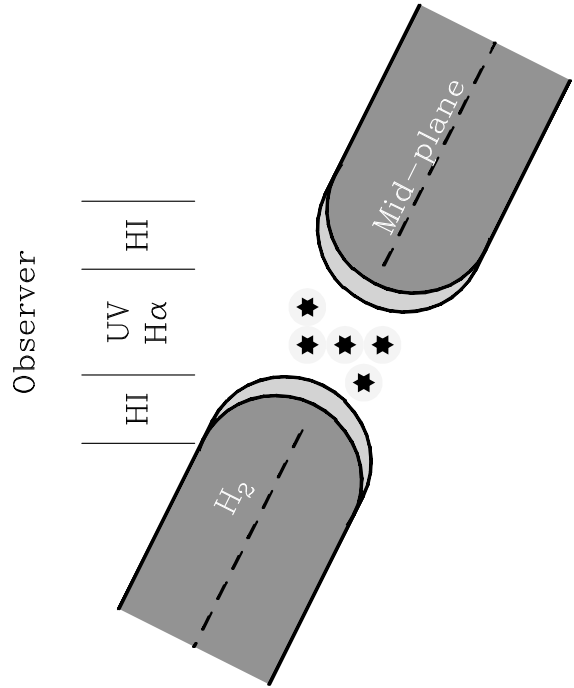
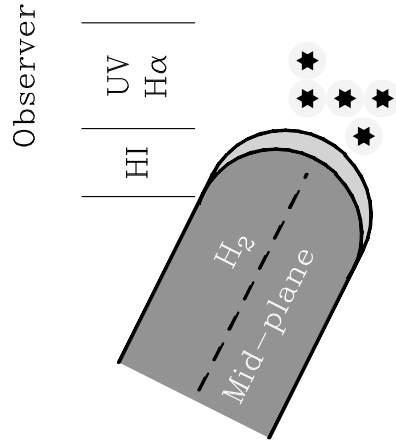
\end{document}